\documentclass[11pt,a4paper]{article}
\usepackage[hyperref]{acl2018}
\usepackage{times}
\usepackage{latexsym}
\usepackage{graphicx}
\usepackage{amssymb}
\usepackage{tipa}
\graphicspath{ {images/} }
\usepackage{svg}
\usepackage{amsmath}
\usepackage{url}

\renewcommand\footnotemark{}

\aclfinalcopy 

\title{Foreign English Accent Adjustment by Learning Phonetic Patterns}

\author{Fedor Kitashov$^\sharp{}^\dagger$\thanks{Work done while at Cisco Systems, San Jose, CA. These authors were supported by funding from Cisco Systems.}, Elizaveta Svitanko$^\sharp{}^\flat$, Debojyoti Dutta$^\sharp$ \\
  $^\sharp$Cisco Systems, USA \\
  $^\dagger$Moscow Institute of Physics and Technology, Russia\\
  $^\flat$Higher School of Economics, Russia\\
  {\tt \{fkitasho, esvitank, dedutta\}@cisco.com} \\
  {\tt \{fedor.kitashov\}@phystech.edu} \\
  {\tt \{eisvitanko\}@edu.hse.ru} \\
  }

\date{}

\begin{document}
\maketitle
\begin{abstract}
State-of-the-art automatic speech recognition (ASR) systems struggle with the lack of data for rare accents. For sufficiently large datasets, neural engines tend to outshine statistical models in most natural language processing problems. However, a speech accent remains a challenge for both approaches. Phonologists manually create general rules describing a speaker's accent, but their results remain underutilized. In this paper, we propose a model that automatically retrieves phonological generalizations from a small dataset. This method leverages the difference in pronunciation between a particular dialect and General American English (GAE) and creates new accented samples of words. The proposed model is able to learn all generalizations that previously were manually obtained by phonologists. We use this statistical method to generate a million phonological variations of words from the CMU Pronouncing Dictionary and train a sequence-to-sequence RNN to recognize accented words with 59\% accuracy.

\end{abstract}

\section{Introduction}


ASR systems have advanced due to increasing dataset sizes and more complex acoustic and language models ~\cite{Dahl:12,  Graves:13, DBLP:journals/corr/BahdanauCSBB15}. Speech processing algorithms perform efficiently on native English speakers' intonation and diction, because researchers can access relatively large datasets with natives' speech samples~\cite{Chan:15}. Recently, Yang et al.~\shortcite{Yang:17} showed that handling speech accents separately leads to the improvement in the word error rate on British and American accents. More non-native accented speech data is necessary to enhance the performance of the existing models. However, its synthesis is still an open problem. In this paper, we propose a method to statistically analyze various accents, automatically extract phonological generalizations and use the trained model to generate accented versions of words. We use the Carnegie Mellon University (CMU) Pronouncing Dictionary~\cite{CMU}. With the proposed approach we increase the size of the CMU dataset from 103,000 phonetic transcriptions with a single accent to one million samples with multiple accents. The other proposed method is that of an accent modification for obtaining better speech recognition performance. Taking phonetic transcriptions as a representation of speech, we can consider accent modification as a spell-checking task. To solve it, we apply sequence-to-sequence recurrent neural network~\cite{Sutskever:14} in order to standardize accented samples.

\begin{table*}
\centering
\begin{tabular}{lll}
  
  \hline
  \\
Original text & Please call Stella. Ask her to bring these things with her ... \\
Manually transcribed sample & p\textsuperscript{h}li\textlengthmark \textsubring{z} k\textsuperscript{h}\textscripta \textlengthmark l\textsuperscript{\textgamma}  st\textepsilon l\textschwa \hspace{0pt}     \textlowering{\ae}sk h\textschwa \hspace{0pt} t\textschwa \hspace{0pt} b\textturnr \~\textsci \textipa{N} \hspace{0pt} \textsubbridge{n}i\textlengthmark \textsubring{z} \texttheta \~\textsci \textrtailn \textsubring{z} w\textsci \texttheta \hspace{0pt} h\textrhookschwa \hspace{0pt} ... \\
Automatically simplified sample & pliz c\textscripta ll st\textepsilon l\textschwa \hspace{0pt}  \ae sk h\textschwa \hspace{0pt} t\textschwa \hspace{0pt}  br\textsci \textipa{N} \hspace{0pt} niz \texttheta \textsci \textrtailn z  w\textsci \texttheta \hspace{0pt} h\textschwa \hspace{0pt} ... \\
\end{tabular}
\caption{The original text, full and simplified phonetic transcriptions
  }
\end{table*}

\section{Dataset Overview}


We take accented speech samples from the Speech Accent Archive built by George Mason University ~\cite{GMU}. Subjects of this research were asked to read a paragraph that contains common English words and practically all English sounds: 
\begin{quote}
``Please call Stella. Ask her to bring these things with her from the store: Six spoons of fresh snow peas, five thick slabs of blue cheese, and maybe a snack for her brother Bob. We also need a small plastic snake and a big toy frog for the kids. She can scoop these things into three red bags, and we will go meet her Wednesday at the train station.''
\end{quote}

 We use a version of the dataset that contains 2511 audio files with 239 unique ethnology codes. Half of the obtained recordings were transcribed using International Phonetic Alphabet (IPA) ~\cite{IPA}. Phonologists manually extract features (phonological generalizations) from this dataset. Each generalization represents the difference in the pronunciation between the particular audio and a GAE sample audio. They may look as follows: the difference between [pliz] and [plis] is the last consonant sound. Voiced sound [z] was turned into its voiceless counterpart [s]. It is called {\em final obstruent devoicing}. However, this type of feature extraction leads to hard-coding manually obtained rules, and it becomes inefficient as the number of generalizations grows. Our goal is to take
 advantage of the huge variety of accents represented in the GMU dataset. We design a model that can automatically generalize such rules based on a small number of samples with unique accents. Earlier, Kunath~\shortcite{Kunath:2009} presented a similar system performing the comparison between two phonetically transcribed texts of human speech. 
 
\subsection{Phonetic Transcription Simplification}


The ``Please call Stella'' dataset was manually transcribed using IPA. In order to provide the most accurate transcriptions, phonologists use special characters to underline the sounds that do not exist in English and the style of pronunciation (e.g. voiceless, aspirated, nasalized etc.). Refer to the Table 1 for the example. 

In this paper, we use the CMU Pronouncing Dictionary as a dataset of natives' pronunciations to generate their accented versions. For more than 100,000 words stored in the CMU dictionary, its phonetic transcriptions consist of only 39 IPA sounds and has only general pronunciation features. We create a reduction dictionary, that maps 169 unique characters from the GMU dataset to 39 unique sounds from the CMU dictionary. Link for the proposed reduction dictionary can be found in References.

\section{Statistical Model}


The statistical model is designed to automatically create phonological generalizations. According to the list of generalizations made in GMU (see the first column in Table 3), it appears that most of them indicate small changes such as swapping two sounds or changing the last sound in the word.
Hence, for each of 69 words from ``Please, call Stella'' we fetch the information of how sounds differ from the ones in GAE and store it to reuse later. More precisely, for every single utterance we find insertions, deletions and replacements which are required to get the corresponding transcription from a GAE sample. For each sound in the dataset we create a dictionary that stores the necessary information about how it is being changed across the dataset. The obtained sound statistics are:

\begin{enumerate}
\item Total number of the sound occurrences in the dataset.
\item Number of times each sound replaced the given one. 
\item Number of times each sound was inserted before or after the given one. 
\item Number of times the given sound was deleted from the GAE pronunciation.
\end{enumerate}

Iterating over pairs ``accented sample -- GAE sample'' we update the number of occurrences for every sound and 4 dictionaries mentioned earlier: replacements, deletions, insertions before and after the given sound. The dictionaries' keys are the sounds being used to modify the GAE pronunciation and the values are the numbers of occurrences for particular changes (see Table 2). Finally, for every sound we compute the probability of being replaced by a particular sound and the probability of being deleted or inserted before or after the given sound. Sounds can be also replaced by multiple sounds and vice versa. For example,
\begin{quote}
pliz --- b\textschwa liz \\
replace: [p] $\leftrightarrow$ [b\textschwa] \\
equal:   \hspace{4.9pt} [liz] $=$ [liz]
\end{quote}

\begin{table}[t!]
\begin{tabular}{|l|p{42mm}|}
\hline 
occurrences & 82 \\ \hline
deletions & 0 \\ \hline
replacements & [e]: 5, [\textschwa]: 3, [e\textschwa]: 1, \newline [\textsci]: 9, [e\textsci]: 1, [\oe]: 1, [\textsci l]: 1 \\ \hline
insertions before & [\textsci]: 1, [\textschwa]: 1, [i]: 1, [j]: 1 \\ \hline
insertions after & [t]: 1, [\textscripta]: 1, [\textschwa]: 1, \newline [d\textschwa]: 1, [d]: 3 \\
\hline
\end{tabular}
\caption{\label{font-table} Statistics obtained for the \textepsilon \hspace{0pt} sound }
\end{table}

\subsection{Automatic Data Generation}

Model learns the probability that particular sound or a sequence of sounds is changed. Iterating through the characters of the input word, it changes particular ones according to the obtained knowledge of possible replacements, insertions and deletions. For example, the word ``milk`` has a phonetic transcription [m\textsci lk]. Our model predictions for [m\textsci lk] are:
\begin{quote}
m\textsci lg, m\textsci lh, n\textsci lk, melk, mi\textschwa lk, m\textepsilon lk, m\textupsilon lk, m\textsci lk, m\textsci \textschwa rk, 
m\textsci rk, m\textsci wk, m\textschwa \textsci lk, miilk, mu\textsci lk, m\textsci \texttheta lk, mil
\end{quote}

Where the most probable modifications are:

\begin{quote}
Insertion: [miilk] \\
Replacement: [n\textsci lk] \\ 
Deletion: [mil] \\ 
\end{quote}
Such an approach allows us to generate more accented versions for every word from the CMU dictionary and therefore get a larger dataset.

\subsection{Comparison With Manually Created Features}

The statistical model was trained on both 
GMU version with 169 unique characters and simplified version with 39 sounds from the CMU Phonological Dictionary. CMU Dictionary contains less sounds, thus some of the very specific pronunciation traits have been lost after the process of data simplification. As a result, for CMU simplified representation the model was able to learn 13 out of 20 generalizations. IPA complex symbols that were used in GMU dataset provide almost full information about one's accent. Trained on this dataset the model is able to learn all phonological generalizations.  

\begin{table}[t!]
\begin{center}
\begin{tabular}{|l|c|c|}
\hline \bf Generalizations & \bf CMU & \bf GMU \\ \hline
final obstruent devoicing & $\checkmark$ & $\checkmark$ \\
consonant voicing &$\checkmark$  & $\checkmark$ \\
stop $\rightarrow$ fricative &   & $\checkmark$ \\
interdental fricative change &$\checkmark$  & $\checkmark$ \\
palatalization &   & $\checkmark$ \\
retroflexing & & $\checkmark$ \\
alveolar approximant change &$\checkmark$  &$\checkmark$\\
w $\rightarrow$ fricative &$\checkmark$ &$\checkmark$ \\
dentalization &$\checkmark$ & $\checkmark$\\
h $\rightarrow$ velar fricative &$\checkmark$ &$\checkmark$ \\
sh $\rightarrow$ s &$\checkmark$ & $\checkmark$\\
stop $\rightarrow$ implosive & &$\checkmark$ \\
labialization &  &$\checkmark$ \\
vowel raising & &$\checkmark$\\
vowel shortening &$\checkmark$ & $\checkmark$ \\
vowel lowering & &$\checkmark$ \\
vowel insertion &$\checkmark$ &$\checkmark$ \\
consonant deletion &$\checkmark$ &$\checkmark$ \\
cluster reduction &$\checkmark$ &$\checkmark$ \\
consonant insertion &$\checkmark$ &$\checkmark$ \\
\hline
\end{tabular}
\end{center}
\caption{\label{font-table} Comparison between manually and automatically obtained phonological generalizations }
\end{table}

\section{Accent Modification}

The aim of this experiment is to adjust the non-English speakers' utterances closer to the correct CMU dictionary versions by learning real-world samples pronunciation specifications. Complex models such as sequence-to-sequence recurrent neural networks (seq2seq RNN) have shown good results on the language processing and modification problems ~\cite{Cho, Ghosh:17}. 
However, these models trained on the samples from ``Please, call Stella'' dataset learn all the patterns of the input sequences and do not obtain a generalization ability. To avoid the problem of overfitting we provide a larger corpus with artificially generated accented phonetic transcriptions using the proposed statistical model.

\subsection{Model Structure}

The model trained on the augmented CMU dataset tries to adapt to the correct transcription version as a spell-checker does. It tries to get rid of unnecessary sounds or change it in an appropriate way so that the model output is as close to the GAE word version as possible. Thus, we choose a seq2seq RNN ~\cite{Cho} to generate the GAE phonetic transcription version character-by-character. The model consists of encoder and decoder parts. The samples with a standard pronunciation from the CMU dictionary represent the decoder input.
The longest word in the corpus contains 34 characters. However, we use 99 percent of the words whose length does not exceed 14 characters. 
Both parts of the system are Long-Short Term Memory (LSTM) layers with 256 neurons. The encoder processes input transcription character-wise and returns an internal state containing the necessary information about the transcription structure. The decoder is trained to predict the next characters of the target sequence, given previous characters of the target sequence. Therefore, the decoder generates the output transcription character-by-character based on the accent patterns which are obtained from the initial state of the encoder.

\subsection{Experiment}


The model has been trained with the batch size of 4096 as the closest maximum which fits in the memory, RMSprop optimizer for the gradient descent, 256 latent dimensions and 100 epochs. 
The input is approximately 800,000 samples with one of the artificially added accents which happens to be Russian. The data was divided into train, validation and test samples in the ratio 80/10/10. We chose accuracy as a quality metric. 

\begin{figure}[ht]
  \centering
  \includegraphics[width=8.5cm, height=6cm]{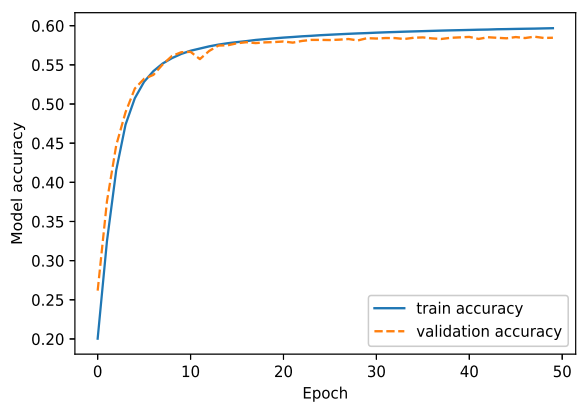}
  \caption{Model accuracy on 50 epochs of training}
\end{figure}

Model shows satisfactory quality on the training set. Starting from 25$^{th}$ epoch the model tends to overfit as the validation accuracy stops increasing along with the training accuracy (Figure 1). 
On the test set the accuracy reaches a point of 0.593. One of the explanations for such a model behaviour is the excess of automatically generated features. This may be considered by the model as a random noise, therefore it is not able to learn the artificial accent patterns properly.

 \nocite{Code}
 \nocite{huang1990hidden}
 \nocite{povey2011kaldi}
 \nocite{CMU}
 \nocite{mohamed2012acoustic}
 \nocite{hinton2012deep}
 \nocite{Graves:13}
 \bibliography{acl2018}
 \bibliographystyle{acl_natbib}



\end{document}